\documentclass[reqno]{amsart}
\usepackage{amsmath}%
\usepackage{amsfonts}%
\usepackage{amssymb}%
\usepackage{graphicx}
\usepackage{setspace}
\usepackage{microtype}

\theoremstyle{plain}

\numberwithin{equation}{section}
\begin{document}
\onehalfspacing
\raggedbottom
\title{An Outline of the Bayesian Decision Theory}
\author{H.R.N.~van~Erp}
\author{R.O.~Linger}


\author{P.H.A.J.M.~van~Gelder}


\begin{abstract}
In this fact sheet we give an outline on the Bayesian Decision Theory. 
\end{abstract}

\maketitle

\section{Introduction}
The Bayesian decision theory is very simple in structure. Its algorithmic steps are the following:
\begin{enumerate}
	\item Use the product and sum rules of Bayesian probability theory to construct outcome probability distributions.
	\item If our outcomes are monetary in nature, then by way of the Bernoulli utility function we may map utilities to the monetary outcomes of our outcome probability distributions.
	\item Maximize the position of the resulting utility probability distributions. 
\end{enumerate}
This is the whole of the Bayesian decision theory.

\section{Constructing Outcome Probability Distributions}
\label{chapter2.1}
In the Bayesian decision theory each problem of choice is understood to consist of a set of decisions from which we must choose. Each possible decision has associated with it its own set of possible outcomes, and each outcome has its own plausibility of occurring relative to the other outcomes under that same decision. Stated differently, each decision in our problem of choice admits its own outcome probability distribution.

In its most abstract form, we have that each problem of choice consists of a set of potential decisions
\[
	D_{i} = \left\{D_{1},\ldots,D_{n} \right\}.
\] 
Each decision $D_{i}$ we make may give rise to a set of possible events 
\[
	E_{j_{i}} = \left\{E_{1_{i}},\ldots, E_{m_{i}} \right\}.
\]
These events $E_{j_{i}}$ are associated with the decisions $D_{i}$ by way of the conditional probabilities $P\!\left(\left.E_{j_{i}}\right|D_{i}\right)$. Furthermore, each event $E_{j_{i}}$ allows for a set of potential outcomes
\[
	O_{k_{j_{i}}} = \left\{O_{1_{j_{i}}},\ldots, O_{l_{j_{i}}}\right\}. 
\]
These outcomes $O_{k_{j_{i}}}$ are associated with the events $E_{j_{i}}$ by way of the conditional probabilities $P\!\left(\left.O_{k_{j_{i}}}\right|E_{j_{i}}\right)$. 

By way of the product rule \cite{Jaynes03}, we compute the bivariate probability distribution of an event and an outcome conditional on the decision taken:
\begin{equation}
	\label{eq.P1.2.6}
	P\!\left(\left.E_{j_{i}}, O_{k_{j_{i}}}\right| D_{i}\right)= P\!\left(\left.E_{j_{i}}\right|D_{i}\right) P\!\left(\left.O_{k_{j_{i}}}\right|E_{j_{i}}\right).
\end{equation}
The outcome probability distribution is then obtained by marginalizing, by way of the sum rule \cite{Jaynes03}, over all the possible events:
\begin{equation}
	\label{eq.P1.2.7}
	P\!\left(\left. O_{k_{j_{i}}}\right| D_{i}\right) = \sum_{j_{i} = 1}^{m_{i}} P\!\left(\left.E_{j_{i}}, O_{k_{j_{i}}}\right| D_{i}\right).
\end{equation}
The outcome probability distributions \eqref{eq.P1.2.7}, for $i = 1,\ldots, n$, are the information carriers which represent our state of knowledge in regards to the objective consequences of our decisions.

\section{A Consistency Proof of the Bernoulli Utility Function}
\label{BernoulliC}
We will now derive the Bernoulli utility function, or, equivalently, the Weber-Fechner law, or, equivalently, in content, Steven's Power law, using the desiderata of invariance and consistency. In this we follow a venerable Bayesian tradition, \cite{Cox46, Jaynes03, Knuth10}.

Say, we have the positive quantities $x$, $y$, and $z$, of some stimulus or commodity of interest. Then these quantities, being numbers on the positive real, admit an ordering. So, let quantities be ordered as $x \leq y \leq z$. We now want to find the function $f$ that quantifies the perceived decrease associated with going from, say, the quantity $z$ to the quantity $x$.

The first functional equation is based on the desideratum that the unknown function $f$ should be invariant for a change of scale in our quantities:
\begin{equation}
	\label{BerC.1}
	f\!\left(x, z\right) = f\!\left(c x, c z\right),
\end{equation}
where $c$ is positive constant. 

For example, if our quantities concern sums of money, then the perceived loss of going from ten dollars to one dollar should be the same perceived loss if we reformulate this scenario in dollar cents.

The second functional equation is based on the desideratum of consistency, in which we state that the perceived decrease in going directly from $z$ to $x$, ought to be the same perceived decrease in going from $z$ to $x$ via $y$:
\begin{equation}
	\label{BerC.2}
	f\!\left(x, z\right) = g\!\left[f\!\left(x, y\right), f\!\left(y, z\right)\right].
\end{equation}

For example, if our quantities concern sums of money, then the perceived loss of going from ten dollars to one dollar should be the same perceived loss if we first go from ten dollars to five dollars, and then from five dollars to one dollar; seeing that in both scenarios we start out with an initial wealth of ten dollars, only to end up with a current wealth of one dollar.

The general solution to \eqref{BerC.1} is \cite{vanErp15a}: 
\begin{equation}
	\label{BerC.1b}
		f\!\left(x, y\right) = h\!\left(\frac{x}{y}\right),
\end{equation}
were $h$ is some arbitrary function.The general solution to \eqref{BerC.2} is \cite{Knuth10}:
\begin{equation}
	\label{BerC.2b}
		\Theta\!\left[f\!\left(x, z\right)\right] = \Theta\!\left[f\!\left(x, y\right)\right] + \Theta\!\left[f\!\left(y, z\right)\right],
\end{equation}
where $\Theta$ is some arbitrary monotonic function. Moreover, because of this arbitrariness, we may define $\Theta$ as \cite{Knuth10}:
\begin{equation}
	\label{BerC.2c}
		\Theta\!\left(x\right) = \log \Psi\!\left(x\right).
\end{equation}
Using \eqref{BerC.2c}, we may rewrite \eqref{BerC.2b}, without any loss of generality, as
\begin{equation}
	\label{BerC.2d}
		\log \Psi\!\left[f\!\left(x, z\right)\right] = \log \Psi\!\left[f\!\left(x, y\right)\right] + \log \Psi\!\left[f\!\left(y, z\right)\right],
\end{equation}
or, equivalently,
\begin{equation}
	\label{BerC.2e}
		\Psi\!\left[f\!\left(x, z\right)\right] = \Psi\!\left[f\!\left(x, y\right)\right] \: \Psi\!\left[f\!\left(y, z\right)\right].
\end{equation} 
Substituting \eqref{BerC.1b} into \eqref{BerC.2b} and \eqref{BerC.2e} and letting, respectively,
\begin{equation}
	\label{BerC.2f}
		\theta\!\left(\frac{x}{y}\right) = \Theta\!\left[h\!\left(\frac{x}{y}\right)\right],
\end{equation} 
and
\begin{equation}
	\label{BerC.2g}
		\psi\!\left(\frac{x}{y}\right) = \Psi\!\left[h\!\left(\frac{x}{y}\right)\right],
\end{equation}  
we obtain the equivalent functional equations:
\begin{equation}
	\label{BerC.2h}
\theta\!\left(\frac{x}{z}\right)  = \theta\!\left(\frac{x}{y}\right) + \theta\!\left(\frac{y}{z}\right)
 \end{equation}
and
\begin{equation}
	\label{BerC.2i}
\psi\!\left(\frac{x}{z}\right)  = \psi\!\left(\frac{x}{y}\right)  \: \psi\!\left(\frac{y}{z}\right) .
 \end{equation}

If we assume differentiability, then \eqref{BerC.2h}, together with the two boundary conditions:
\begin{equation}
	\label{BerC.3}
	f\!\left(x, x\right) = \theta\!\left(\frac{x}{x}\right) = 0,
\end{equation}
and
\begin{equation}
	\label{BerC.4}
	f\!\left(x, y\right) = \theta\!\left(\frac{x}{y}\right) < 0,  \quad \text{for } x < y,
\end{equation}
is sufficient to find the function $f$ that quantifies the perceived decrease associated with going from the quantity $y$ to the quantity $x$. 

This function $\theta$ turns out to be Bernoulli's utility function, or, equivalently, the Weber-Fechner law of sense perception:
\begin{equation}
	\label{BerC.4b}
	f\!\left(x, y\right) = q \log \frac{x}{y},		\qquad q \geq 0
\end{equation}
where $y$ is our initial asset position and $x$ is the final asset position, and $q$ is some arbitrary constant which has to be obtained by way psychological experimentation. 

So, Bernoulli's utility function \eqref{BerC.4b} is the only function that adheres to the desiderata of unit invariance and consistency, respectively, \eqref{BerC.1} and \eqref{BerC.2}, and the boundary conditions that a zero change should lead to a zero perceived loss and that a perceived loss should be assigned a negative value, respectively, \eqref{BerC.3} and \eqref{BerC.4}. Any other utility function will be in violation with these fundamental desiderata and specific boundary conditions. 

Note that Fechner re-derived \eqref{BerC.4b} in 1860 as the law that guides our sensory perception. In the years that followed \eqref{BerC.4b} proved to be so successful, as it, amongst other things, gave rise to our decibel scale, that it established psychology as a legitimate experimental science \cite{Fancher90}. But as Fechner was very careful, for metaphysical reasons, or so we hazard to guess \cite{vanErp15a}, to apply his Weber law, which later became the Fechner-Weber law, only to non-monetary stimuli, the implied universality of \eqref{BerC.4b} was not recognized for the longest time. 

However, because of the here given consistency derivation of \eqref{BerC.4b}, it is now shown that the Fechner-Weber, or, equivalently, Bernoulli's utility function, is one of the consistent functions that quantifies the distance between $x$ and $y$; thus, explaining the universal applicability of Bernoulli's utility function. 

The other consistent distance function is Steven's power law, which may be derived as follows. If we assume differentiability, then \eqref{BerC.2i}, together with the two boundary conditions:
\begin{equation}
	\label{BerC.5}
	f\!\left(x, x\right) = \psi\!\left(\frac{x}{x}\right) = 1,
\end{equation}
and
\begin{equation}
	\label{BerC.6}
	0 < f\!\left(x, y\right) = \psi\!\left(\frac{x}{y}\right) < 1,  \quad \text{for } x < y,
\end{equation}
is sufficient to find the function $f$ that quantifies the perceived decrease associated with going from the quantity $y$ to the quantity $x$. 

This function $f$ turns out to be Steven's power law: 
\begin{equation}	
	\label{BerC.6b}
	f\!\left(x, y\right) = \left(\frac{x}{y}\right)^{q},		\qquad q \geq 0.
\end{equation}
where $y$ is our initial asset position and $x$ is the final asset position, and $q$ is some arbitrary constant which has to be obtained by way psychological experimentation. 

So, Steven's power law \eqref{BerC.6b} is the only function that adheres to the desiderata of unit invariance and consistency, respectively, \eqref{BerC.1} and \eqref{BerC.2}, and the boundary conditions that a zero change should lead to a ratio of one between the initial and final `asset position' and that a perceived loss should be assigned a value smaller than 1, respectively, \eqref{BerC.5} and \eqref{BerC.6}. Any other utility function will be in violation with these fundamental desiderata and specific boundary conditions.  

We summarize, given the desiderata \eqref{BerC.1} and \eqref{BerC.2}, the Fechner-Weber law \eqref{BerC.4b} results from the boundary condition that negative increments result negative utilities and a zero increment results in an utility of zero, \eqref{BerC.3} and \eqref{BerC.4}; whereas Steven's power law \eqref{BerC.6b} results from the boundary condition that utilities must be greater than zero and that a zero increment results an utility of one, \eqref{BerC.5} and \eqref{BerC.6}. 

Stated differently, the Fechner-Weber law and Steven's power law are both equivalent in content, differing only in the proposed utility scale. A subtlety that seems to have been overlooked by some, seeing that the Fechner-Weber law versus the Steven's power law has been a source of controversy in psycho-physical community\cite{Stevens61}.

\section{The Criterion Of Choice as a Degree of Freedom}
\label{Choice}


Let $D_{1}$ and $D_{2}$ be two decisions we have to choose from. Let $o_{i}$, for $i = 1, \dots, n$, and $o_{j}$, for $j = 1, \dots, m$, be the monetary outcomes associated with, respectively, decisions $D_{1}$ and $D_{2}$. Then in the Bayesian decision theory we first construct the two outcome distributions that correspond with these decisions:
\begin{equation}
	\label{eq.why.1}
	p\!\left(\left.o_{i}\right|D_{1}\right), \qquad  p\!\left(\left.o_{j}\right|D_{2}\right),
\end{equation}
where, if $n = m$, the outcomes $o_{i}$ and $o_{j}$ may or not may be equal for $i = j$.

We then proceed, by way of the Bernoulli utility function \eqref{BerC.4b}, or, equivalently, the Weber-Fechner law, to map utilities to the monetary outcomes $o_{i}$ and $o_{j}$ in \eqref{eq.why.1}. This leaves us with the utility probability distributions:
\begin{equation}
	\label{eq.why.2}
	p\!\left(\left.u_{i}\right|D_{1}\right), \qquad  p\!\left(\left.u_{j}\right|D_{2}\right).
\end{equation}

Now, our most primitive intuition regarding the utility probability distributions \eqref{eq.why.2} is that the decision which corresponds with the utility probability distribution which lies more to the right will also be the decision that promises to be the most advantageous. So, when making a decision we ought to compare the positions of the utility probability distributions on the utility axis and then choose that decision which maximizes the position of these utility probability distributions. 

This all sounds intuitive enough. But how do we define the position of a probability distribution? Ideally we would have some consistency derivation of what constitutes a position measure of a probability distribution, say,
\begin{equation}
	\label{eq.why.2b}
		H_{n}\!\left(p_{1}, \ldots, p_{n}, x_{1}, \ldots, x_{n}\right)
\end{equation}
where $p_{i}$ are the probabilities of the values $x_{i}$, for $i = 1, \ldots, n$. But in the absence of such a consistency derivation we have to take our recourse to \textit{ad hoc} common sense considerations. Stated differently, the criterion of choice in our decision theory constitutes a degree of freedom. 

\subsection{The Expectation Value as a Position Measure}
From the introduction of expected outcome theory in the 17th century and expected utility theory in the 18th century the implicit assumption has been that the measure of a position of a probability distribution is given by its expectation value \cite{Jaynes03, Bernoulli38}:
\begin{equation}
	\label{eq.why.2c}
		E\!\left(X\right) = \sum_{i=1}^{n} p_{i} x_{i} = H_{n}\!\left(p_{1}, \ldots, p_{n}, x_{1}, \ldots, x_{n}.\right)
\end{equation}
But this criterion of choice has proven to be so unsatisfactory that it has given rise to the paradigm of behavioral economics which holds as its central tenet that human decision making does not adhere to the maximization of expectation values \cite{Kahneman11}. So, we set out to search for a more appropriate criterion of choice.    

\subsection{The Confidence Bounds as a Position Measure}
Now we may imagine a decision problem in which we are only interested in the positions of the probabilistic worst or best case scenarios. 

The absolute worst case scenario is:
\begin{equation}
	\label{eq.why.2d}
		a = \min\!\left(x_{1}, \ldots, x_{n}\right). 
\end{equation}
The criterion of choice \eqref{eq.why.2d} is also known as the maximin criterion of choice. 

The k-sigma lower bound of a given probability distribution is a given as
\begin{equation}
	\label{eq.why.2e}
		L\!B\!\left(X\right) = E\!\left(X\right) - k \; \text{std}\!\left(X\right), 
\end{equation}
where
\begin{equation}
	\label{eq.why.2f}
		\text{std}\!\left(X\right) = \sqrt{\sum_{i=1}^{n} p_{i} \left[x_{i} - E\!\left(X\right)\right]^{2}}, 
\end{equation}
and where $k$ is the sigma level of the lower bound. The probabilistic worst case scenario then is given as:
\begin{equation}
	\label{eq.why.2g}
		L\!B^{*}\!\left(X\right) = \begin{cases} E\!\left(X\right) - k \; \text{std}\!\left(X\right), \qquad &L\!B\!\left(X\right) > a, \\
		a, \qquad &L\!B\!\left(X\right) \leq a.
		\end{cases}
\end{equation}
So, we have that the probabilistic worst case scenario holds the maximin criterion of choice as a special case.

The absolute best case scenario is:
\begin{equation}
	\label{eq.why.2h}
		b = \max\!\left(x_{1}, \ldots, x_{n}\right). 
\end{equation}
The criterion of choice \eqref{eq.why.2h} is also known as the maximax criterion of choice. 

The k-sigma upper bound of a given probability distribution is a given as:
\begin{equation}
	\label{eq.why.2i}
		U\!B\!\left(X\right) = E\!\left(X\right) + k \; \text{std}\!\left(X\right), 
\end{equation}
where $k$ is the sigma level of the upper bound. The probabilistic best case scenario then is given as:
\begin{equation}
	\label{eq.why.2j}
		U\!B^{*}\!\left(X\right) = \begin{cases} E\!\left(X\right) + k \; \text{std}\!\left(X\right), \qquad &U\!B\!\left(X\right) < b, \\
		b, \qquad &U\!B\!\left(X\right) \geq b.
		\end{cases}
\end{equation}
So, we have that the probabilistic best case scenario holds the maximax criterion of choice as a special case.

If we take as our criterion of choice \eqref{eq.why.2g} then we only endeavor to minimize our `losses' and if we take as our criterion of choice \eqref{eq.why.2j} then we only endeavor to maximize our `gains'. A more rational, that is, balanced, criterion of choice would be to make a trade-off between the losses/gains in the probabilistic worst case scenarios \eqref{eq.why.2g} and the corresponding gains/losses in the probabilistic best case scenarios \eqref{eq.why.2j}.

\subsection{The Sum of Confidence Bounds as a Position Measure}
If we take as our criterion of choice
\begin{equation}
	\label{eq.why.2k}
		\frac{L\!B^{*}\!\left(X\right) + U\!B^{*}\!\left(X\right)}{2} = \begin{cases} E\!\left(X\right), \qquad &L\!B\!\left(X\right) > a, U\!B\!\left(X\right) < b, \\
		\frac{a + E\!\left(X\right) + k \; \text{std}\!\left(X\right)}{2} , \qquad &L\!B\!\left(X\right) \leq a, U\!B\!\left(X\right) < b, \\
		\frac{E\!\left(X\right) - k \; \text{std}\!\left(X\right) + b}{2}, \qquad &L\!B\!\left(X\right) > a, U\!B\!\left(X\right) \geq b, \\
				\frac{a + b}{2}, \qquad &L\!B\!\left(X\right) \leq a, U\!B\!\left(X\right) \geq b,
		\end{cases}
\end{equation}
then we have a position measure which makes a trade-off between the losses/gains in the probabilistic worst case scenarios \eqref{eq.why.2g} and the corresponding gains/losses in the probabilistic best case scenarios \eqref{eq.why.2j}; see Appendix A. 

This alternative position measure, as an added benefit, also holds the traditional criterion of choice \eqref{eq.why.2c} as a special case, when no undershoot and overshoot of, respectively, the lower and upper sigma confidence bounds occur, as well as Hurwitz's criterion of choice with a balanced pessimism factor of $c = 1/2$, when both an undershoot and an overshoot occur. Nonetheless, it may be found that the criterion of choice \eqref{eq.why.2k} is vulnerable to a simple counter-example. 

Imagine two utility probability distributions having equal lower and upper bounds, but one being right-skewed and the other being left-skewed. Then the criterion of choice \eqref{eq.why.2k} will leave us undecided between the two, whereas our intuition would give preference to the decision corresponding with the left-skewed distribution, as the bulk of the probability distribution of the left-skewed distribution will be more to the right than that of the right-skewed distribution. 

\subsection{The Sum of Confidence Bounds Plus the Expectation Value as a Position Measure}
What we seek to maximize in our decision theory is the position of the utility probability distributions; as we have that the decision that puts our utility probability distribution most to the right promises to be the most profitable decision. In this there is little room for maneuvering. But in our choice of the measure that captures the position of a given probability distribution there is all the more. 

Taking a cue from the behavioral economists we have derived as an alternative to \eqref{eq.why.2c} the criterion of choice \eqref{eq.why.2k} that also takes into account the standard deviation of a given probability distributions, by way of the positions of the under and overshoot corrected lower and upper bounds. But only to find its universality compromised by the simple counter example of a right-skewed and a left-skewed distribution which have the same lower and upper bounds.   

Now, also taking a cue from the intuitive results which flow forth from \eqref{eq.why.2k} \cite{vanErp15a}, we may `repair' our criterion of choice \eqref{eq.why.2k}, albeit in an \textit{ad hoc} fashion, by taking as our position measure for a probability distribution the weighted sum:
\begin{equation}
	\label{eq.why.16}
	\frac{L\!B^{*}\!\left(u\right) + E\!\left(u\right) + U\!B^{*}\!\left(u\right)}{3} = \begin{cases} E\!\left(X\right), \qquad &L\!B\!\left(X\right) > a, U\!B\!\left(X\right) < b, \\
		\frac{a + 2 E\!\left(X\right) + k \; \text{std}\!\left(X\right)}{3} , \qquad &L\!B\!\left(X\right) \leq a, U\!B\!\left(X\right) < b, \\
		\frac{2 E\!\left(X\right) - k \; \text{std}\!\left(X\right) + b}{3}, \qquad &L\!B\!\left(X\right) > a, U\!B\!\left(X\right) \geq b, \\
				\frac{a + E\!\left(X\right) + b}{3}, \qquad &L\!B\!\left(X\right) \leq a, U\!B\!\left(X\right) \geq b,
		\end{cases}
\end{equation}
For in this criterion of choice we not only take into account the trade-off between the probabilistic worst and best case scenarios, but also the location of the bulk of the probability density in a uni-model probability distribution; thus, accommodating the intuitive preference for the left-skewed distribution of the above counter example.

The position measure \eqref{eq.why.16} is the weighted sum of the positions of, respectively, the probabilistic worst, expected, and best case. The uncorrected lower and upper bounds, \eqref{eq.why.2e} and \eqref{eq.why.2i}, have been traditionally used as simplifying proxies for their generating probability distributions, by way of confidence intervals:
\begin{equation}
	\label{eq.why.19}
\left[L\!B\!\left(X\right), U\!B\!\left(X\right)\right].
\end{equation}
We, alternatively, take as our simplifying proxy the corrected lower and upper bounds, \eqref{eq.why.2g} and \eqref{eq.why.2j}, and the expectation value \eqref{eq.why.2c}: 
\begin{equation}
	\label{eq.why.20}
\left[L\!B^{*}\!\left(X\right), E\!\left(X\right), U\!B^{*}\!\left(X\right)\right].
\end{equation}
Because of the corrections for lower bound undershoot and upper bound overshoot in \eqref{eq.why.20} we have that for skewed distributions the distance between $E\!\left(X\right)$ and $L\!B^{*}\!\left(X\right)$ may differ from the distance between $U\!B^{*}\!\left(X\right)$ and $E\!\left(X\right)$; thus, reflecting the asymmetry present in these distributions. 

The position of the generating probability distribution then is taken to be the weighted sum of the positions of the elements of our simple proxy distribution. This then is the rationale behind the criterion of choice \eqref{eq.why.16}.
  
\section{Discussion} 
It may be read in Jaynes' \cite{Jaynes03}, that to the best of his knowledge, there are as of yet no formal principles at all for assigning numerical values to loss functions; not even when the criterion is purely economic, because the utility of money remains ill-defined. In the absence of these formal principles, Jaynes final verdict was that decision theory can not be fundamental. 

The Bernoulli utility function, initially derived by Bernoulli, by way of common sense first principles \cite{Bernoulli38}, has now been derived by way of a consistency argument. This consistency argument explains why it is that Bernoulli's utility function, both in its original Fechner-Weber law and in its alternative Steven's power law form, has proven to be so ubiquitous and successful the field of sensory perception research; simply because human sense perception, like the laws of Nature, adheres to the desideratum of consistency. 

%

The history of Bayesian probability theory has taught us that the usefulness of a theory, in terms of its practical and beautifully intuitive results, in the absence of a compelling axiomatic basis, provides no safeguard against attacks by those who choose to close their eyes to this usefulness. This is why we felt compelled to search for a consistency derivation of the Bernoulli utility function. 

Now, having presented a consistency proof for the Bernoulli utility function, the question now is: Is the Bayesian decision theory, just like the Bayesian probability and information theories, Bayesian in the strictest sense in the word, or, equivalently, an inescapable consequence of the desideratum of consistency? We will now try to answer this question.

The first two algorithmic steps of the Bayesian decision theory, respectively, the construction of outcome probability distributions by way of the Bayesian probability theory and the construction of utility probability distributions by way of the Bernoulli utility function, allow us no freedom. 

To construct our outcome and utility probability distributions otherwise, would be to invite inconsistency. But there is one degree of freedom remaining in the Bayesian decision theory as a whole. This remaining degree of freedom lies in the choice of our position measure of a given probability distribution. 

In any problem of choice we will endeavor to choose that decision which has a corresponding utility probability distribution that is lying most the right on the utility axis; that is, we will choose to maximize our utility probability distributions. In this there is little freedom. But we are free, in principle, to choose the measures of the positions of our utility probability distributions any way we see fit. Nonetheless, we believe that it is always a good policy to take into account all the pertinent information we have. 

If we only maximize the expectation values of the utility probability distributions, then we will, by definition, neglect the information that the standard deviations of the utility probability distributions have to bear on our problem of choice, by way of the symmetry breaking in the case of an overshoot of one of the bounds. 

Likewise, we are free to only maximize one of the confidence bounds of our utility probability distributions, while neglecting the other. But in doing so, we will be performing probabilistic maximin or maximax analyses, and, consequently, neglect the possibility of either the (catastrophic) losses in the lower bound or the (astronomical) gains in the upper bound. 

However, if we only maximize the sum of the lower and upper bound, or a scalar multiple  thereof, then we will make a trade-off between the probabilistic worst and best case scenarios. But in the process, we will, for uni-modal distributions, be neglecting the location of the bulk of our probability distributions. 

This is why, in our minds, the scalar multiple the sum of the lower bound, expectation value, and upper bound currently is the most all-round position measure for a given probability distribution, as it reflects the position of the probabilistic worst and best case scenarios, as well as the position of the expected outcome.      

Having removed the degree of freedom of the utility function by way of a consistency derivation, we now should endeavor to find a like consistency derivation of the measure of the position of a given probability distribution \eqref{eq.why.2b}; as such a consistency derivation would make the Bayesian decision theory incontestable. 

But until that time, we will have to do with the kind of simplistic common sense reasoning that led us from the traditional position measure \eqref{eq.why.2c}, to the position measures \eqref{eq.why.2k} and \eqref{eq.why.16}, and make the disclaimer that our adopted criterion of choice, as a degree of freedom, is just a matter of choice.
\\
\\
\noindent\textbf{Acknowledgments:} We would like here to express our gratitude to Kevin H. Knuth, whose kind and patient feedback led us to our consistency proof of Bernoulli's utility function and to Kevin M. Vanslette, whose simple but highly effective `counter' example led us to drop \eqref{eq.why.2k} and propose \eqref{eq.why.16} instead. The research leading to these results has received partial funding from the European Commission's Seventh Framework Program [FP7/2007-2013] under grant agreement no.265138.


%
%
%

\appendix
\section{Deriving the Sum of the Lower and Upper Confidence Bound Measure}
Now, the confidence bounds of \eqref{eq.why.2}, say:
\begin{equation}
	\label{eq.why.3}
	\left[L\!B\!\left(\left.u\right|D_{1}\right), U\!B\!\left(\left.u\right|D_{1}\right)\right], \qquad  \left[L\!B\!\left(\left.u\right|D_{2}\right), U\!B\!\left(\left.u\right|D_{2}\right)\right],
\end{equation}
may provide us with a numerical handle on the concept of more-to-the-right.

For example, if we have that both
\begin{equation}
	\label{eq.why.4}
	L\!B\!\left(\left.u\right|D_{1}\right) > L\!B\!\left(\left.u\right|D_{2}\right),  \qquad U\!B\!\left(\left.u\right|D_{1}\right) > U\!B\!\left(\left.u\right|D_{2}\right).
\end{equation}
Then we will have an unambiguous preference for decision $D_{1}$ over decision $D_{2}$; seeing that under both the still probable worst and best case we will be better if we opt for $D_{1}$.

Likewise, if we have that either
\begin{equation}
	\label{eq.why.5}
	L\!B\!\left(\left.u\right|D_{1}\right) = L\!B\!\left(\left.u\right|D_{2}\right),  \qquad U\!B\!\left(\left.u\right|D_{1}\right) > U\!B\!\left(\left.u\right|D_{2}\right),
\end{equation}
or
\begin{equation}
	\label{eq.why.6}
	L\!B\!\left(\left.u\right|D_{1}\right) > L\!B\!\left(\left.u\right|D_{2}\right),  \qquad U\!B\!\left(\left.u\right|D_{1}\right) = U\!B\!\left(\left.u\right|D_{2}\right).
\end{equation}
Then, again, we will have an unambiguous preference for decision $D_{1}$ over decision $D_{2}$. In the constellation \eqref{eq.why.5}, we stand, all other things being equal, to be better of under the still probable best case scenario; while in the constellation \eqref{eq.why.6}, we stand, all other things being equal, to be less worse of under the still probable worst case scenario.

However, things become more ambiguous when, say, under decision $D_{1}$, we have to make a trade-off between either a gain in the upper bound and a loss in the lower bound
\begin{equation}
	\label{eq.why.7}
	L\!B\!\left(\left.u\right|D_{1}\right) < L\!B\!\left(\left.u\right|D_{2}\right), \qquad U\!B\!\left(\left.u\right|D_{1}\right) > U\!B\!\left(\left.u\right|D_{2}\right),
\end{equation}
or a gain in the lower bound and a loss in the upper bound
\begin{equation}
	\label{eq.why.8}
	L\!B\!\left(\left.u\right|D_{1}\right) > L\!B\!\left(\left.u\right|D_{2}\right), \qquad U\!B\!\left(\left.u\right|D_{1}\right) < U\!B\!\left(\left.u\right|D_{2}\right).
\end{equation}

We \textit{postulate} here that a rational criterion of choice in the respective trade-off situations \eqref{eq.why.7} and \eqref{eq.why.8}, would be to pick that decision whose gain in either the lower or upper bound exceeds the loss in the corresponding upper or lower bound. 

So, if, say, under $D_{1}$ we stand to gain more in the still probable best case scenario than we stand to lose under the still probable worst case scenario, that is, \eqref{eq.why.7}:
\begin{equation}
	\label{eq.why.9}
	  L\!B\!\left(\left.u\right|D_{2}\right) - L\!B\!\left(\left.u\right|D_{1}\right) <  U\!B\!\left(\left.u\right|D_{1}\right) - U\!B\!\left(\left.u\right|D_{2}\right),
\end{equation}
then we will choose $D_{1}$ over $D_{2}$. Likewise, if under $D_{1}$ we stand to gain more in the still probable worst case scenario than we stand to lose under the still probable best case scenario, that is, \eqref{eq.why.8}:
\begin{equation}
	\label{eq.why.10}
	  L\!B\!\left(\left.u\right|D_{1}\right) - L\!B\!\left(\left.u\right|D_{2}\right) >  U\!B\!\left(\left.u\right|D_{2}\right) - U\!B\!\left(\left.u\right|D_{1}\right),
\end{equation}
then again we will choose $D_{1}$ over $D_{2}$.

Note that the gains and losses in this discussion pertain to gains and losses on the utility dimension, not on the monetary outcome dimension. On the utility dimension the phenomenon of loss aversion, that is, the phenomenon that monetary losses may weigh heavier than equal monetary gains, has already been accounted for. Stated differently, the utility scale is a linear loss-aversion corrected scale for the moral value of monies.

Now, if we look at the scenarios \eqref{eq.why.7} and \eqref{eq.why.8}, and the corresponding postulated rational, because intuitive, criteria of choice \eqref{eq.why.9} and \eqref{eq.why.10}, then we see that we will choose $D_{1}$ over $D_{2}$ whenever we have that
\begin{equation}
	\label{eq.why.11}
	L\!B\!\left(\left.u\right|D_{1}\right) + U\!B\!\left(\left.u\right|D_{1}\right) > L\!B\!\left(\left.u\right|D_{2}\right) + U\!B\!\left(\left.u\right|D_{2}\right).
\end{equation}
Moreover, this single criterion of choice is also consistent with the choosing of $D_{1}$ over $D_{2}$ in the scenarios \eqref{eq.why.4}, \eqref{eq.why.5}, and \eqref{eq.why.6}. 


Note that if the decision inequality \eqref{eq.why.11} goes to an equality:
\begin{equation}
	\label{eq.why.12}
	L\!B\!\left(\left.u\right|D_{1}\right) + U\!B\!\left(\left.u\right|D_{1}\right) = L\!B\!\left(\left.u\right|D_{2}\right) + U\!B\!\left(\left.u\right|D_{2}\right).
\end{equation}
Then we have that we will be undecided when it comes to the decisions $D_{1}$ and $D_{2}$. 

Now, the $k$-sigma bounds in \eqref{eq.why.3} translate to
\begin{equation}
	\label{eq.why.13}
	\left[E\!\left(\left.u\right|D_{i}\right) - k \: \text{std}\!\left(\left.u\right|D_{i}\right), E\!\left(\left.u\right|D_{i}\right) + k \: \text{std}\!\left(\left.u\right|D_{i}\right)\right], 
\end{equation}
for $i = 1, 2$, which, if substituted in \eqref{eq.why.11}, give the inequality
\begin{equation}
	\label{eq.why.15}
	2 E\!\left(\left.u\right|D_{1}\right) > 2 E\!\left(\left.u\right|D_{2}\right),
\end{equation}
which brings us right back to Bernoulli's expected utility theory, as proposed in 1738, in which it is proposed that we choose that decision which maximizes the expectation value of the utility probability distributions \cite{Bernoulli38}. 

Nonetheless, the criterion of choice, that the sum of the upper and lower bound should be maximized, as proposed here, will deviate from Bernoulli's initial 1738 proposal when the $k$-sigma intervals overshoot either the minimal or the maximal value of the utility probability distributions, or both.

Let $a$ and $b$, respectively, be the minimal and maximal values of a given utility probability distribution. Then we may identify three additional cases, relative to \eqref{eq.why.15}:
\begin{equation}
	\label{eq.why.15b}
	L\! B^{*}\!\left(u\right) + U\! B^{*}\!\left(u\right) = 
	\begin{cases} 2 \: E\!\left(u\right), \qquad  &L\! B\!\left(u\right) \geq a, \quad U\! B\!\left(u\right) \leq b \\
	a + U\! B\!\left(u\right), \qquad &L\! B\!\left(u\right) < a, \quad U\! B\!\left(u\right) \leq b \\
	L\! B\!\left(u\right) + b, \qquad   &L\! B\!\left(u\right) \geq a, \quad U\! B\!\left(u\right) > b \\
	a + b, \qquad   &L\! B\!\left(u\right) < a, \quad U\! B\!\left(u\right) > b
	\end{cases}
\end{equation}
These additional cases correspond, respectively, to scenarios where the $k$-sigma intervals \eqref{eq.why.13} either undershoot, or overshoot, or both undershoot and overshoot the minimal and maximal values of a given utility probability distribution. 

In the case of a lower confidence bound undershoot (e.g cases two and four) a too pessimistic worst case scenario is in play, and in the case of an upper confidence bound overshoot a too optimistic a best case scenario is in play, which is why we have to readjust these confidence bounds by replacing them with more realistic worst and best case scenarios (e.g the minimal and maximal values, $a$ and $b$, of a given utility probability distribution).  

It may be readily seen that any scalar multiple of \eqref{eq.why.15b} will retain the transitive ordering of the criterion of choice \eqref{eq.why.15b}. Now, if we take as our scalar multiple $c = 1/2$, then our criterion of choice may be interpreted as a location measure of utility probability distribution:
\begin{equation}
	\label{eq.why.15c}
	\frac{L\! B^{*}\!\left(u\right) + U\! B^{*}\!\left(u\right)}{2} = 
	\begin{cases} E\!\left(u\right), \qquad  &L\! B\!\left(u\right) \geq a, \quad U\! B\!\left(u\right) \leq b \\
	\frac{a + U\! B\!\left(u\right)}{2}, \qquad &L\! B\!\left(u\right) < a, \quad U\! B\!\left(u\right) \leq b \\
	\frac{L\! B\!\left(u\right) + b}{2}, \qquad   &L\! B\!\left(u\right) \geq a, \quad U\! B\!\left(u\right) > b \\
	\frac{a + b}{2}, \qquad   &L\! B\!\left(u\right) < a, \quad U\! B\!\left(u\right) > b
	\end{cases}
\end{equation}
where we note that the first case of \eqref{eq.why.15c} corresponds with Bernoulli's expected utility theory criterion of choice, whereas the fourth case corresponds with Hurwitz's criterion of choice with a balanced pessimism coefficient of $\alpha = 1/2$.  

Note also that we have an added degree of freedom in our decision theory in that we may put an explicit premium on either caution or opportunity. For example, in the absence of lower bound undershoot and upper bound overshoot, we have as the lower and upper bounds whose sum is to be maximized:
\begin{equation}
	\label{eq.why.16q}
	L\!B^{*}\!\left(u\right) = E\!\left(u\right) - k_{1} \: \text{std}\!\left(u\right),
\end{equation}	
and
\begin{equation}
	\label{eq.why.17}
	U\!B^{*}\!\left(u\right) = E\!\left(u\right) + k_{2} \: \text{std}\!\left(u\right).
\end{equation}
Then \eqref{eq.why.16q} and \eqref{eq.why.17} sum to:
\begin{equation}
	\label{eq.why.18}
	L\!B^{*}\!\left(u\right) + U\!B^{*}\!\left(u\right) = 2 \: E\!\left(u\right) + \left(k_{2} - k_{1}\right) \: \text{std}\!\left(u\right).
\end{equation}
If in \eqref{eq.why.18} we let $k_{1} > k_{2}$, then we put a premium caution; alternatively, if we set $k_{2} > k_{1}$, then we put a premium on opportunity; and if we let $k_{1} = k_{2}$, then we have an equal trade-off between caution and opportunity taking, \eqref{eq.why.15b}.  

\end{document}